\documentclass[epsfig, 12pt, onecolumn]{article}
\usepackage{epsfig}
\usepackage{amssymb,amsmath}
\usepackage{psfrag}
\newcommand{\newc}{\newcommand}
\newc{\gsim}{\lower.7ex\hbox{$\;\stackrel{\textstyle>}{\sim}\;$}}
\newc{\lsim}{\lower.7ex\hbox{$\;\stackrel{\textstyle<}{\sim}\;$}}
\newc{\gev}{\,{\rm GeV}}
\newc{\mev}{\,{\rm MeV}}
\newc{\ev}{\,{\rm eV}}
\newc{\kev}{\,{\rm keV}}
\newc{\tev}{\,{\rm TeV}}
\def\ln{\mathop{\rm ln}}

\newc{\mz}{M_Z}
\newc{\mpl}{M_{\rm pl}}
\newc{\mw}{m_{\rm weak}}
\newc{\nr}[1]{N^c_R{}_{#1}}

\def\beq{\begin{equation}}
\def\eeq{\end{equation}}
\def\bea{\begin{eqnarray}}
\def\eea{\end{eqnarray}}
\def\bitem{\begin{itemize}}
\def\eitem{\end{itemize}}
\newc{\ie}{{\it i.e.}}          \newc{\etal}{{\it et al.}}
\newc{\eg}{{\it e.g.}}          \newc{\etc}{{\it etc.}}
\newc{\cf}{{\it c.f.}}
\def\bar#1{\overline{#1}}
\def\vev#1{\left\langle #1 \right\rangle}

\def\inv{^{\raise.15ex\hbox{${\scriptscriptstyle -}$}\kern-.05em 1}}
\def\lbar{{\lower.35ex\hbox{$\mathchar'26$}\mkern-10mu\lambda}} 

\def\to{\rightarrow}

\let\p=\partial
\let\<=\langle
\let\>=\rangle

\let\+=\uparrow
\let\al=\alpha

\let\ep=\epsilon

\let\la=\lambda
\let\La=\Lambda

\let\si=\sigma

\let\th=\theta

\begin{document}
\thispagestyle{empty}
\vspace*{.5cm}
\noindent
\hspace*{\fill}{\large  OUTP-06 16 P}\\
\vspace*{2.0cm}

\begin{center}
{\Large\bf Warped Axions}
\\[2.5cm]
{\large Thomas Flacke $^a$, Ben Gripaios $^{a,b}$, John March-Russell $^a$, and~David~Maybury $^a$
}\\[.5cm]
{\it $^a$ Rudolf Peierls Centre for Theoretical Physics\\
University of Oxford, 1 Keble Road, Oxford OX1 3NP, UK}
\\[.2cm]
{\it $^b$ Merton College, Oxford OX1 4JD, UK}
\\[.2cm]
(Dated: November 22, 2006)
\\[1.1cm]

{\bf Abstract}\end{center}
\noindent
We study a number of realizations of axions existing in a multi-`throat' generalization of the warped throat geometry
of a Randall-Sundrum slice of $AdS_5$.  As argued by previous authors, the problem of generating a suitable, phenomenologically allowed Peccei-Quinn scale is simply and elegantly solved by the warping.   In compactifications with two or more throats it is possible to simultaneously solve the Standard Model hierarchy problem by the Randall-Sundrum mechanism while implementing interesting warped axion models.  The constructions discussed are related to holographic duals of previously studied models of composite axions arising from strongly coupled four-dimensional dynamics.
\newpage

\setcounter{page}{1}

\section{Introduction}

One of the outstanding puzzles of the Standard Model (SM) of particle physics is the strong
CP problem \cite{Peccei:2006as}.  Within the SM there exist two independent sources of CP-violation; the
phase $\phi$ in the CKM quark mixing matrix, and the CP-violating term 
\beq
S_{\th} = \frac{\bar{\th}}{64 \pi^2} \ep_{\mu\nu\la\si} F^{a\mu\nu} F^{a\la\si}  
\eeq
associated with the strong interactions where
\beq
\bar{\th} = \th -\arg\det M_q ,
\label{thetabar}
\eeq
is the physical $U(1)_{\rm chiral}$-invariant coupling
(here $\th$ is the bare QCD CP-violating theta-parameter, and $M_q$ is the primordial quark
mass matrix).  The magnitudes of the two CP-violating parameters are quite different: while the CKM phase
$\phi$ must be ${\cal O}(1)$ to explain the observed CP violation in the $K$- and $B$-meson systems, the
strong interaction CP-violating parameter is constrained to be $\bar{\th} \lsim 10^{-9}$ by the non-observation
of a neutron electric dipole moment \cite{rpp}.  This disparity is particularly puzzling given the fact that the large
CKM phase would naively indicate that $\arg\det M_q$ is also large, and thus there must be an almost perfect cancellation
in Eq.(\ref{thetabar}).  The extreme smallness of $\bar{\th}$ has no explanation in the SM.  

This presents a major opportunity for beyond-the-standard-model physics. 
Probably the most promising idea is that of Peccei and Quinn \cite{pq} whereby the SM is enhanced by
a global $U(1)_{PQ}$ symmetry, broken spontaneously at a scale $f_{PQ}$, and explicitly by a
$U(1)_{PQ} SU(3)_c^2$ anomaly, implying \cite{ww} that there
exists a new pseudoscalar degree of freedom, the axion, $a$, which couples to QCD via the term
\beq
\Delta S = \frac{a}{32\pi^2 f_{PQ}} F\tilde{F} .  
\eeq
Strong QCD dynamics then generates a potential for $a$ such that $\bar{\th}$ is naturally relaxed to zero
by the vacuum expectation value of $a$.  

The properties of the axion are highly constrained by laboratory, astrophysical, and cosmological
considerations.  The original Peccei-Quinn-Weinberg-Wilczek implementation was quickly
excluded by direct collider searches, leading to the construction of a number of so-called `invisible'
axion models \cite{ksvzaxion,dfszaxion} where the PQ-breaking scale $f_{PQ}$ is much greater than the electroweak
scale.  Later analysis of stellar and red giant cooling constraints led to the lower bound
of $f_{PQ}\gsim 10^{9} \gev$ \cite{rpp}.  There is also an upper bound $f_{PQ}\lsim 10^{11}\gev$ arising from the
requirement that axion dark matter produced by the misalignment mechanism does not overclose the universe \cite{rpp}.

This upper bound on the PQ scale poses problems for axions in string theory.  Although axion-like
states are ubiquitous in string theory compactifications \cite{stringaxions}, the effective PQ scale is  
generically near the string scale, thus strongly violating the cosmological closure bound on $f_{PQ}$.
(One exception to this statement are the large-volume flux-stabilized models
of Ref.\cite{conlon} in which it appears to be possible to suppress the string scale and thus string axion $f_{PQ}$
scale down to $\sim 10^{11}\gev$.  However these models have other phenomenological issues, \eg,
gauge-coupling unification.)

In this letter we propose to use the warped geometry of the throats that commonly occur in string constructions
with flux \cite{throats}  to address this difficulty.  The problem of generating a suitable PQ scale is simply and elegantly
solved by the warping as in the Randall-Sundrum solution to the hierarchy problem \cite{RS}.   Moreover, as
recently argued by a number of authors \cite{savas,csaki,hmr},
multi-throat geometries with a range of warp factors are not un-natural in string compactifications, and here
we further argue that in compactifications with two or more throats, it is possible to simultaneously solve
the SM hierarchy problem by the Randall-Sundrum mechanism while implementing
such warped axion models with phenomenologically allowed couplings. 

Finally we wish to mention some precursor papers to this study.  In particular Dienes, Dudas and Gherghetta
\cite{DDG} considered the phenomenology of axions and their KK excitations in a flat bulk, Choi \cite{Choi}
discussed a warped one-throat model of the axion solving the $f_{PQ}$ problem based upon a bulk $U(1)$
gauge theory broken by boundary conditions, and Collins and Holman \cite{Collins:2002kp} considered
a warped solution to the $f_{PQ}$ problem for axions arising from a complex scalar field in one throat model
with a third mobile brane.  In this letter we extend these works discussing the detailed implementation
and structure of such models in a two-throat geometry.\footnote{After completion of this paper we were
kindly informed of another study \cite{KimKim}, in the context of string compactifications, of the
possibility of creating a $f_{PQ}/M_{Pl}$ hierarchy using warping and/or mixing between multiple
string axions.  The constructions differ from the models presented here.} 
 
\section{Two Throats}\label{twothroats}

Our constructions are based on a warped 5d geometry with two-throats -- one containing the
SM, in particular the Higgs and SM fermion degrees of freedom, while the other
throat provides the axion.  The geometry of both throats will be approximated
by the standard Randall-Sundrum slice of anti-deSitter (AdS) space with metric
\beq
ds^2 = (k z)^{-2}(\eta_{\mu\nu} dx^\mu dx^\nu - d^2 z ),
\label{metricz}
\eeq
where their common curvature radius is $1/k$.  For the first throat the coordinate lies in the
range $z\in [1/k,L_1]$, while the second has coordinate $w$ with range $w\in [1/k,L_2]$.
The two throats are glued together at a `UV brane' at position
$z=1/k$ in the first throat and $w=1/k$ in the second, so these coordinate values are
identified. The two throats are terminated by `IR branes' at $z=L_1$ and $w=L_2$.  We will call
these the PQ brane and SM brane respectively, although often we will take the 
SM gauge fields, but not the Higgs or SM fermions, to propagate in both throats.\footnote{Sometimes
it will be more convenient to work
in terms of the 5th dimensional coordinate $y$ defined by $z=\exp(k y)/k_1$ where
$y\in [0,y_{PQ}]$ with associated metric
$ ds^2 = \exp(-2k y)\eta_{\mu\nu} dx^\mu dx^\nu - d^2 y$ 
in which the exponential warping is made manifest $L_1 = \exp(k y_{PQ})/k$, and 
similarly for the second throat with obvious replacements. We note for completeness
that it is possible that the curvature radius differs in the two throats, in which case
there are modifications to some aspects of the phenomenology, but not of the fundamental
mechanisms that we discuss in later sections.}

A simple analysis of the graviton zero mode localized about the UV brane shows
that the 4D effective Planck mass is given in terms of the underlying 5d Planck mass $M_5$ 
by the Randall-Sundrum like formula
\beq
\mpl^2 =  \frac{2 M_5^3}{k}.
\label{planck}
\eeq
up to exponentially small corrections. The factor of 2 is due to the presence of the two
throats.  In this letter we assume the usual Randall-Sundrum choice of $k\sim M_5/10$, and
thus $M_5 \sim \mpl/{\rm few}$ so that the derivative expansion is under control but
no small dimensionless parameters appear in the fundamental Lagrangian.

\section{Axions from Bulk Complex Scalars}

First, consider a simple model with a free complex bulk scalar field $\Phi$ in the PQ
throat with action
\beq
S_\Phi = M_5 \int d^5x \sqrt{-G} \left( G^{MN} \p_M \Phi^* \p_N \Phi - m^2 |\Phi|^2 \right) 
\label{bulkS}
\eeq
(here $M,N=0,1,2,3,5$).  We decompose $\Phi = \eta \exp(i a)$ and in addition impose a
Dirichlet boundary condition for $\eta$ at the PQ brane,
and Neumann conditions for $\eta$ at the UV brane and for $a$ at both branes:
\bea
\eta|_{IR} = v & \qquad & \p_{z}\eta |_{UV} = 0 \\
\p_{z} a |_{IR} = 0 & \qquad & \p_{z} a |_{UV} = 0 ,
\label{bc}
\eea
(these conditions are consistent with a well-posed variational problem). 
The IR boundary condition for $\eta$ which forces it to acquire an expectation
value can be thought of arising from
the decoupling limit of an IR-brane-localized interaction of the form
$\la (|\Phi|^2 - v^2)^2$ in a way similar to some implementations of
Higgs-less electroweak
symmetry breaking models \cite{csaki}.  We take the two dimensionless
parameters $m/M_5$ and $v/M_5$ both to be $O(1)$.  

Writing $\eta(x,z)= \exp(-ip\cdot x) \eta(z)$, the solution to the $\eta$
equation of motion
\beq
\p_z \left[ \left(\frac{1}{k z}\right)^3 \p_z \eta \right] - \frac{m^2}{k^5 z^5} \eta
= -\frac{p^2}{k^3 z^3} \eta 
\label{etaeqn}
\eeq
for $p^2=0$, subject to the boundary conditions, gives the $z$-dependent profile of the
$\eta$ expectation value:
\beq
\vev{\eta} =  A z^{2-\nu} + B z^{2+\nu}
\label{etasol}
\eeq
where the exponent $\nu = (4 + m^2/k^2)^{1/2}\geq 0$.  In this letter we will
focus on the case $\nu\geq 2$ (we will return the case $\nu<2$ which has some interesting
features in a later work).  In the limit of significant warping in the PQ throat
$k L_1 \gg 1$, and for generic values of $\nu\neq 2$, the constants are given to
good approximation by
\bea
A & = & v \frac{\nu+2}{\nu-2} \frac{1}{k^{2\nu} L_1^{\nu+2}} , \label{etaconstant1}\\
B & = & v \frac{1}{L_1^{\nu+2}} .\label{etaconstants}
\eea
For the special case $\nu=2$, the solution becomes a constant
\beq
\vev{\eta} = v .
\eeq
Note that for $\nu\neq 2$ the $\eta$ vev is maximized at the PQ brane and 
drops towards the UV brane -- this will have significant phenomenological
implications. 

The equation for the modes of the phase field $a(x,z)=\exp(-ip\cdot x) a(z)$ in the $\vev{\eta}$
background is
\beq
\p_z \left[\vev{\eta(z)}^2 \left(\frac{1}{k z}\right)^3 \p_z a(z) \right] 
= -\vev{\eta(z)}^2 \frac{p^2}{k^3 z^3} a(z). 
\label{aeqn}
\eeq
The boundary conditions, Eq.(\ref{bc}), allow for all $\nu$ a zero mode ($p^2=0$) solution $a_0$
with constant profile in $z$, $a_0=c$.  This Nambu-Goldstone mode will become
the pseudo-Nambu-Goldstone axion once we couple to QCD. 

By solving Eq.(\ref{aeqn}) for non-zero $p^2$, the KK modes of $a(x,z)$ are
found to have $z$-dependent wavefunctions (for $\nu\neq 2$ and $L_1 k \gg 1$) of the
approximate form
\beq
a_n(z) = d_n z^{-\nu} \biggl( Y_\nu(m_n z) J_{1+\nu}(m_n/k) - J_\nu(m_n z) Y_{1+\nu}(m_n/k) \biggr) .
\label{kkmodes}
\eeq
Enforcing the Neumann boundary conditions at both the UV and the IR brane implies that the KK masses $m_n$ are given approximately by
\beq
m_n = \frac{\al_{1+\nu}^{(n)}}{L_1} ,
\label{kkmasses}
\eeq
where $\al_{1+\nu}^{(n)}$ are the zeroes of $J_{1+\nu}(x)$, 
and thus are of order the exponentially warped PQ brane scale, as expected.
A similar analysis of the non-zero modes of Eq.(\ref{etaeqn}) describing the
KK excitations of the radial $\eta$ field is straightforward and yields masses
given by the zeroes of $J_{\nu}(m_n L_1)$, which are thus also of order $1/L_1$.

For $a_0$ to be the strong-CP-solving axion we must couple $a(x,z)$, and thus $\Phi$ in the correct way to the SM.  There are
various possibilities for this coupling which we now discuss in turn.

\subsection{UV brane localized coupling}

If the full SM is restricted to the 2nd throat, and the axion is only in the first throat as
considered above, then the coupling to QCD can only arise from UV-brane localized couplings.
Moreover if the Higgs(es) are localized on the SM brane as in the normal RS construction, then direct
$\Phi^2$-Higgs-Higgs couplings as in the DFSZ model are not possible.\footnote{We note in passing that
following Ref.\cite{bulkhiggs} it is possible to consider models with the SM Higgs having some small amplitude in the bulk of the SM throat.  In this case a phenomenologically viable axion model via a $\Phi^2 H H$ coupling on the UV brane is possible.  Similar constructions may also be possible
in the case of a composite Higgs arising from a bulk gauge field \cite{Contino:2003ve}. We do not
consider such models in detail here.} 
However one can adapt the KSVZ construction \cite{ksvzaxion} of the invisible
axion by introducing coloured fermions on the UV brane with interactions
\beq
S_Q = \int d^4 x \sqrt{ -G_{\rm ind}|_{UV} }
\left( \Phi {\bar Q}_L Q_R + \Phi^* {\bar Q}_R Q_L \right) .
\label{Qaction} 
\eeq
By a U(1) chiral rotation of the $Q$ fields, the axion can be moved from this term to a UV brane
localized interaction with the gluon topological term 
\beq
\int d^4 x \frac{1}{32\pi^2 f_{PQ}} a(x,1/k) F {\tilde F},
\label{topolterm}
\eeq
resulting from the anomalous Jacobian of the $Q$ field-redefinition.  
In the limit that the 
non-zero-mode KK excitations of the axion are much heavier than the QCD scale, the axion is almost
entirely made up of the zero mode $a_0$.  To determine the effective PQ scale $f_{PQ}$ we must
canonically normalize the kinetic term for $a_0$
\beq
c^2 M_5 \int_{1/k}^{L_1} dz \left(\frac{1}{k z}\right)^3 \vev{\eta(z)}^2 = \frac{1}{2}
\eeq
implying
\beq
c = kL_1\sqrt{(1+\nu)k/(M_5  v^2)} .
\eeq
The physical axion is $\bar a = a_0/c$.  This together with Eqs.(\ref{Qaction}-\ref{topolterm})
leads to an expression for the PQ scale:
\beq
f_{PQ} =  \frac{1}{kL_1} \sqrt{\frac{M_5 v^2}{(1+\nu)k}}\sim\frac{1}{kL_1}\frac{\mpl}{\sqrt{2(1+\nu)}}
\label{pqscale}
\eeq 
where, for the last equation we assumed the natural value $v\sim M_5$ and used the Planck mass relation, Eq.(\ref{planck}). 
If we demand $f_{PQ}\sim 10^{11}\gev$ then we find that a
warp factor at the PQ brane of $k L_1 \sim 10^{7}$ is necessary.  Using the Goldberger-Wise mechanism \cite{GW} (or its
generalization in the context of string theoretic throats \cite{BHT}) such a warp factor is both natural and not unusual
despite all underlying dimensionless parameters being $O(1)$.

Unfortunately the states $Q,{\bar Q}$ cause difficulties for the simple model above model.
The expectation value $\vev{\eta}$ in Eqs.(\ref{etasol}-\ref{etaconstants}) together with
UV-brane interaction Eq.(\ref{Qaction})
implies that the exotic quark states have small masses of order
\beq
m_Q \simeq v \left(\frac{1}{k L_1}\right)^{2+\nu}  .
\label{Qmass}
\eeq
This relation immediately rules out models with $\nu >2$, where Eqs. (\ref{etasol}-\ref{etaconstants}) apply, as it implies that
$m_Q < 10^{-6}\gev$, even for the most favorable case of $f_{PQ}\sim 10^{12}\gev$, 
in gross disagreement with the direct collider search limit on coloured fermion states of approximately
$200\gev$, as well as astrophysical and cosmological considerations.

\begin{figure}
\begin{center}
\psfrag{y}[tl]{\parbox[r]{50pt}{\phantom{1} \vspace{80pt}\hspace{-80pt}$\ln(kL_1)$}}
\psfrag{x}[tl]{\parbox[r]{50pt}{\phantom{1} \vspace{20pt}\hspace{-150pt}$\nu-2$}}
\includegraphics{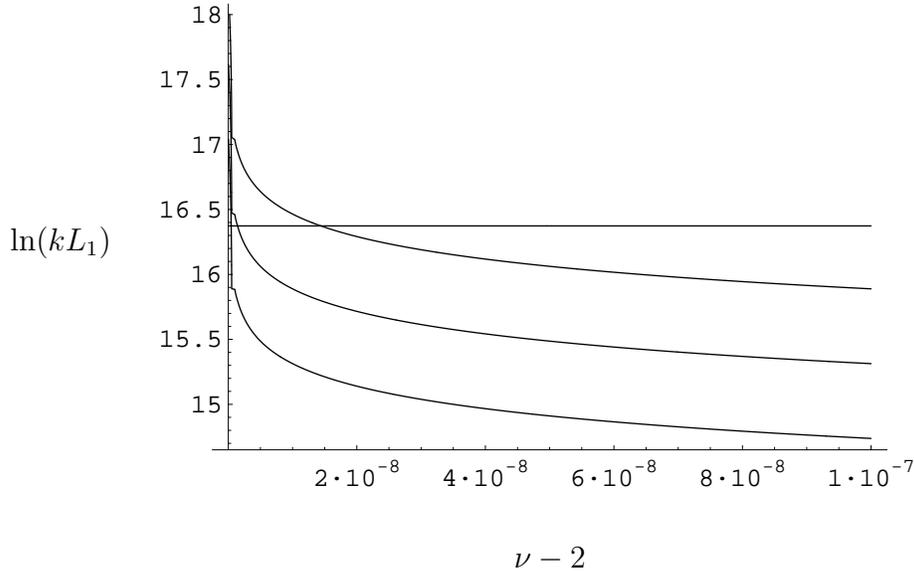}
\vspace{50pt}
\caption{The constraint that $f_{PQ}= 10^{12} \gev$ (almost constant) along with lines giving $m_Q$ ($100 \gev$, $1 \tev$,
and $10 \tev$ reading from top to bottom)
as a function of $\ep = \nu-2$ near zero on the $x$-axis and $\ln(k L_1)$ on the $y$-axis.}
\label{nearepzero}
\end{center}
\end{figure}

The limit of $\nu\simeq 2$ provides a phenomenologically allowed region, as the
profile of $\vev{\eta}$ is nearly flat implying that the $Q$'s are heavier as the vacuum expectation value
on the UV brane is less suppressed.   However this only occurs for $\nu$ extremely close to 2, where the approximate relations in Eqs. (\ref{etaconstant1}-\ref{etaconstants}) do not hold. Defining
$\ep=\nu-2$, Fig.\ref{nearepzero}  shows the result of an exact numerical evaluation of $m_Q$
as a function of $\ep$ and $\ln(k L_1)$. Masses significantly greater than $1\tev$ consistent with
$f_{PQ}\sim 10^{12}\gev$ require $\ep \lsim 10^{-8}$.  Such extreme fine-tuning implies that there
should exist a symmetry reason for taking $\nu =2$ (\ie, $m_\Phi = 0$). However, an analysis of the limit $\ep \rightarrow 0$ shows, that at $\ep=0$, $f_{PQ}$ cannot be warped down sufficiently unless an extreme fine-tuning is imposed, rendering this particular setup un-natural.

\subsection{PQ-brane localized couplings}\label{PQlocalized}

A less constrained arrangement is to have $Q,{\bar Q}$ localized 
on the PQ brane but $\Phi$ still in the bulk.  Since $Q,{\bar Q}$ are colour-charged,
this requires that at least the $SU(3)_c$ gauge fields propagate in the bulk
of both throats.\footnote{In fact if we are to have unification of gauge couplings then
it is preferred to have an $SU(5)$ or $SO(10)$ bulk unified gauge theory in both throats with
boundary conditions on the branes that break the unified group down to the SM.  It is straightforward to
construct such models along the lines developed for one-throat Randall-Sundrum theories \cite{gaugeunif}.}

As shown in Ref.\cite{csaki} the matching boundary conditions at the UV brane
for the $A_\mu$ components in $N$ different throats is (in the decoupling limit and
in unitary gauge, and assuming for simplicity equal AdS curvature
radii and gauge couplings):
\bea
& A^a_{(i)\mu} = A^a_{(i+1)\mu} \\
& \sum_{i=1}^N  \frac{1}{k z_i} \p_{z_i} A^a_{(i)\mu} =0 .
\eea
Imposing these conditions on the $SU(3)$ colour gauge fields in the two throats, and 
requiring Neumann boundary conditions at the IR branes leads to solutions which include a
massless physical 4d zero mode with constant profile in both throats. 

Since the axion still arises from a bulk complex scalar field the effective PQ breaking scale
remains warped down as given in Eq.(\ref{pqscale}).  However as the $Q,{\bar Q}$ exotic quarks are localized to the PQ brane, $\vev{\eta}$, immediately leads to a mass of order
\beq
m_Q \sim \frac{v}{k L_1}= f_{PQ} \sqrt{\frac{k}{M_5(1+\nu)}}  .
\label{Qmass2}
\eeq
For the usual Randall-Sundrum choice of parameters, $k\sim M_5/10$, this leads to
$m_Q \sim f_{PQ}/3 \gsim 10^{10} \gev$, posing no problem for direct searches or cosmological bounds if
the scale of inflation is sufficiently low.  Thus this simple arrangement of fields in a bulk
yields an attractive solution to the PQ scale problem. 

It is clear that an even more minimal model can be constructed by taking the fields
$\Phi$ and $Q,{\bar Q}$ to be localized on the PQ brane.   
The terms in the action involving the new fields are
\beq
S=\int d^4 x\sqrt{-G_{PQ}} \biggl( \p_\mu \Phi^* \p^\mu \Phi -
\la(|\Phi|^2 - v^2)^2 - \left( \Phi {\bar Q}_L Q_R + \Phi^* {\bar Q}_R Q_L \right) +  \mathcal{L}_{kin}(Q_L,Q_R)\biggr)
\eeq
Again, because the coloured exotic quarks live on the PQ brane, $SU(3)_c$ and possibly
other SM gauge fields must live in the PQ throat.   Now, trivially, both the PQ scale and exotic
quark masses are warped down to the scale set by the IR of the PQ throat which can be
taken to be $10^{10} - 10^{11}\gev$ with no difficulty.

\section{Axions from Bulk Gauge Fields}

In 5D spacetime, the 5th component of a $U(1)$ gauge field $A_M$ provides an attractive 
alternate realization of the axion \cite{Choi}. As a minimal model, consider an abelian gauge
field in the bulk of PQ throat with action
\beq
S= \int d^5 x \sqrt{-g}\, \biggl( -\frac{1}{4g_5^2}\, g^{MN}g^{PQ} F_{MP}F_{NQ}
  -\frac{1}{2\xi g_5^2} \biggl[ g^{\mu\nu} \p_\mu A_\nu 
    + z \xi g^{55} \p_5 \left(\frac{A_5}{z}\right) \biggr]^2 \biggr) .
\label{gaugeS}
\eeq
The term in square brackets is a gauge-fixing term which eliminates the
mixing between $A_5$ and $A_\mu$.  Consistent with a well-defined variational
principle, one can choose boundary conditions for the 
fields such that $A_5$ contains a zero mode, but the massless mode for
$A_\mu$ is projected out (thus leaving no surviving 4D gauge symmetry). Specifically,
we take $A_\mu$ to satisfy Dirichlet conditions on both UV and IR branes while
$A_5$ satisfies (warped) Neumann boundary conditions  
\beq
A_\mu\biggr|_{\rm bdy}  = 0, \qquad \p_5 \left(\frac{A_5}{z}\right)\biggr|_{\rm bdy} = 0 .
\label{gaugebc}
\eeq

The equations of motion arising from the action Eq.(\ref{gaugeS}) are
\bea
\eta^{\mu\nu}\p_\mu\p_\nu A_5 + \xi \p_5\left(z \p_5 \left(\frac{A_5}{z}\right)\right) = 0 \\
\eta^{\mu\si}\eta^{\la\nu}\left( \p_\si F_{\mu\la} + \frac{1}{\xi} \p_\la \p_\mu A_\si\right) + kz\p_5\left(\frac{\eta^{\mu\nu}}{kz} \p_5 A_\mu\right)=0.
\label{gaugeeom}
\eea
In unitary gauge, $\xi\to \infty$, the solutions of the equations of motion
have a particularly simple form: $A_5$ only contains
a zero mode, the KK modes of $A_5$ are eaten by the massive excitations of $A_\mu$, and
the massless $A_\mu$ mode is eliminated by the boundary conditions.

The masses and $z$-dependent profiles of the massive $A_\mu$ modes follow from the equation 
\beq
kz\p_5\left(\frac{\eta^{\mu\nu}}{kz} \p_5 f(z) \right)  - m^2 f(z)=0
\eeq
with solutions, after imposing the UV boundary condition, of the form
\beq
f_n(z) = z\left(-\frac{Y_1(m_n/k)}{J_1(m_n/k)} J_1(m_n z) + Y_1(m_n z)\right) .
\eeq
To a good approximation in the large warp factor limit, the masses are given by the zeroes 
of $J_1(m_n L_1)$, namely, $m_n=\al_1^{(n)}/L_1$.

The $A_5$ zero mode contains the axion degree of freedom. From the equations of motion
it has the form  
\beq
A_5(x,z)  =   N z a(x)  ,
\label{A5zm}
\eeq
where $N$ is a normalization constant.  This zero mode is peaked towards the IR brane. 
Upon substitution of this solution in the kinetic term
of the bulk action Eq.(\ref{gaugeS}), the requirement that $a(x)$ is a canonically
normalized scalar in 4d leads to $N\approx \sqrt{2kg_5^2}/L_1$ in the limit of large warping.
  
Note that the $\mu$-components of the boundary conditions Eq.(\ref{gaugebc}) imply that the 5D gauge transformation parameter $\La(x,z)$ is, at the boundaries, a function of $z$ only, while the $z$-component of the boundary condition implies that $\p_5\La \sim z$ at the boundaries. Therefore after imposing the boundary conditions and working in 5D unitary gauge, a {\it global} subgroup of the original $U(1)_X$ 5D gauge symmetry survives, where the gauge transformation parameter has the fixed functional form
\beq
\La(x,z) = (C z^2+D)  ,
\eeq
where $C$ and $D$ are constants.  Under this transformation $A_5$ has a shift symmetry as befits a Nambu-Goldstone mode.  In order to interpret $A_5$ as the axion, it has to be correctly coupled to the
QCD anomaly.\footnote{Outside of unitary gauge, the natural object which encodes the axion degree of freedom is the Wilson line integral $U(x)=\exp(i\int dz A_5/g_5)$ from the UV to the IR brane, and the couplings for $a(x)$ discussed in the next subsections can all be re-written in terms of the combination
$-iU^{-1}\p_\mu U$.  For simplicity we focus on the unitary-gauge formulation, although the manifestly
non-local nature of $U$ shows that the axion dynamics is protected from possible (local) quantum
gravity violations of the PQ shift symmetry\cite{Choi,quantgrav,AHCCR}.}

\subsection{UV or IR localized couplings}

Taking QCD to be localized in the SM throat, $A_5$ can couple to the QCD anomaly only via a UV-brane
localized interaction of the form
\beq
S_{int} = \int d^4 x \frac{\alpha}{64\pi^2 M_5} A_5 \Big{|}_{\rm UV} \ep^{\mu\nu\la\si}G^a_{\mu\nu} G^a_{\la\si},
\label{locS}
\eeq
where $\alpha$ is a dimensionless parameter, $M_5$ is the characteristic mass scale on the UV brane and $A_5$ is given in unitary gauge by Eq.(\ref{A5zm}).  (The interaction Eq.(\ref{locS}) could, for example, be induced by putting a pair of exotic quarks $Q_L,Q_R$ with suitable PQ charges on the UV brane.)  In terms of the canonically normalized field $a(x)$, the 4D effective action is
\beq
S_{4Deff} = \int d^4 x \left( \frac{1}{2}\p_\mu a(x) \p^\mu a(x)+\frac{1}{64\pi^2 f_{PQ}} a(x) \ep^{\mu\nu\la\si}G^a_{\mu\nu} G^a_{\la\si} \right)
\eeq
where 
\beq
f_{PQ}=\frac{M_5 k L_1}{\sqrt{2k g_5^2}} \sim M_5^2 L_1 \gg M_5 \sim \mpl.
\eeq
Here we have taken the large warping limit, $kL_1\gg 1$, all dimensionful parameters in the 5D action (i.e. $k,g_5^2$) of the order of $M_5$, and the dimensionless coupling $\alpha$ to be of order 1. Thus, in this case the Peccei-Quinn scale is warped up due to the `volume suppression' of the brane localized interaction.  Since this fails to achieve our objective we turn to other arrangements.
 
As in section \ref{PQlocalized}, we now assume QCD additionally propagates in the PQ-throat with the boundary conditions chosen such that the QCD gauge fields possess a zero mode.  We assume the existence of a pair of exotic quarks $Q_L,Q_R$ on the IR brane of the PQ throat charged under $SU(3)_C$ and transforming non-trivially under $U(1)_{PQ}$.  After performing a chiral rotation the only possible form of the interaction between $a(x)$ and QCD is again Eq.(\ref{locS}) but now evaluated at the PQ throat IR brane.
In this case we find 
\beq
f_{PQ}=\frac{M_5}{NL_1}  \sim \sqrt{\frac{M_5^3}{k}} \sim \mpl
\eeq
where we have used the expressions for the normalization constant $N$ and for the 4d Planck mass
Eq.(\ref{planck}). Here again, we fail to reduce the PQ scale. 

\subsection{Bulk interaction}

A successful model of the axion arising from a bulk gauge field requires a bulk interaction between $A_5$
and QCD.  To achieve this, one may again introduce exotic quark states, this time in the bulk.  After
integrating out these exotic quark states, a Chern-Simons interaction of form 
\beq
S_{int} = \int d^5 x \frac{\alpha}{64\pi^2 M_5} \ep^{MNPQR} G^a_{MN} G^a_{PQ}  A_R
\label{bulkint}
\eeq
is induced in the bulk.
 
Using the form Eq.(\ref{A5zm}) of the $A_5$ zero mode, we find the expression for the PQ scale to be
\beq
f_{PQ}=\sqrt{\frac{2}{k g_5^2}} \frac{1}{L_1} \simeq \frac{\mpl}{M_5 L_1},
\label{bulkfpq}
\eeq
where we have used the 4d Planck mass formula Eq.(\ref{planck}) and $g_5^2\sim 1/M_5$.
For the standard choice of $k\sim M_5/10$, we see that $f_{PQ}$ is warped down to the
IR brane scale $1/L_1$.  Note that it is possible to reduce the scale $1/L_1$ while keeping
$f_{PQ}$ constant by increasing the hierarchy $\mpl/M_5$, but this is limited to the
range $1/L_1^2 > f_{PQ}^3/\mpl$ by the requirement that $L_1>1/k$.   Numerically
this allows a IR PQ-throat scale as small as $10^6\gev$.

\section{Holographic Interpretation} 

The bulk mechanisms outlined in the previous sections can be given a purely 4D interpretation
by applying the by-now-well-known holographic rules, mapping a 5D theory in a slice of AdS$_5$
to a broken CFT in 4D, with the fifth-dimensional co-ordinate playing the role of the renormalization group
scale \cite{Arkani-Hamed:2000ds,Rattazzi:2000hs,Perez-Victoria:2001pa}.
The UV and IR branes in 5d correspond, respectively, to cutting off the CFT in the UV at energy scale $k$
(and thereby making the sources of the CFT, given by the values of the bulk fields on the UV brane, dynamical),
and to spontaneous breaking of the CFT in the IR through strong-coupling effects.
A model with multiple throats is dual to a 4-d theory with multiple CFTs, each with its own strong-coupling scale,
and interacting only via external sources.

The dual interpretation of a bulk Abelian gauge field is particularly interesting \cite{Contino:2003ve}.
In a full AdS space, as opposed to just a slice, gauge symmetries in the bulk are dual to global symmetries
of the 4-d theory. The interpretation when the branes are added  depends on the choice of boundary conditions 
for the  $A_{\mu}$ components on the two branes. (The boundary conditions for $A_5$ are necessarily opposite.)
If a Neumann condition is chosen for $A_{\mu}$ on the UV brane, then the source (which is dual to $A_{\mu}(z=1/k)$)
is non-vanishing, and the interpretation is that the global symmetry has been gauged by the dynamical source.
If, in addition, a Neumann condition is chosen on the IR brane, there is a zero mode in the 4-d spectrum and we conclude that the
gauge symmetry persists at low energies. On the other hand, a Dirichlet condition on the IR brane implies no zero mode, 
and we conclude that the gauge symmetry is spontaneously broken in the IR. 

If, by contrast the UV boundary condition is taken to be Dirichlet, then the symmetry is not present in the UV.
In general, it will be explicitly broken by the UV dynamics. The same is true if we impose a Dirichlet condition on the IR brane
except that now we interpret the symmetry breaking as spontaneous, for the following reason.
Dirichlet boundary conditions for $A_{\mu}$ imply Neumann boundary conditions for $A_5$. There is thus an $A_5$ zero mode in the 4-d spectrum, which is
a scalar from the 4-d viewpoint. We interpret this mode, which is massless at tree-level, as the Goldstone boson
that results from the spontaneous breaking of the global symmetry. The mode is localized towards the IR brane, meaning that 
it is interpreted as a composite state in the dual CFT. It acquires a small mass via loop effects that feel the explicit 
symmetry breaking on the UV brane. It is thus a pseudo-Goldstone boson in general.

A bulk Abelian gauge field in a warped geometry as outlined in the previous section thus provides a natural solution of the strong CP problem, where we identify the bulk
$U(1)$ gauge symmetry with the PQ symmetry. The PQ scale (which is the scale at which the $U(1)$ is spontaneously broken)
is given by the IR brane scale. Thus, on the 5-d side, the PQ scale allowed by astrophysical and cosmological constraints is
naturally explained by warping, while on the 4-d side it is explained as the scale at which the CFT coupling becomes strong.

The only other necessary requirements are to endow the $U(1)$ with a colour anomaly, such that the strong CP problem
can be solved and to explain why the PQ scale is very different from the warped-down scale that is relevant for
electroweak symmetry breaking. The former is achieved either by adding exotic quarks with both bulk U(1) and colour charges,
or directly by the 5-d Chern-Simons term. The latter is achieved by putting the U(1) gauge field in the bulk of a different
throat to the Higgs. 

This interpretation of the multi-throat models as strongly-coupled 4-d models has an intriguing parallel with older
composite axion models \cite{compositeaxion} that sought a natural explanation for the PQ scale of the invisible axion via strong-coupling effects. The models presented here provide a concrete, and moreover calculable, realization of these ideas.

\section{Conclusions}

In this paper we have shown how a warped two throat Randall-Sundrum set-up simply and elegantly solves the problem of generating the correct scale of spontaneous PQ-breaking.  A variety of models are possible depending upon the fundamental degree of freedom that contains the axion, and the localization (or otherwise) of the interaction of the axion with QCD.  In section 3 we discussed axions arising from a complex scalar field, either
propagating in the bulk, or localized on the IR brane of the Peccei-Quinn throat, and we coupled
the axion degree of freedom to QCD by employing a variant of the 4-dimensional KSVZ-model construction by introducing
exotic quarks transforming under $U(1)_{PQ}$.  We show that
depending upon their location in the 5d  geometry these exotic states can either be supermassive, or located quite close in mass to the electroweak scale.  A simple and manifestly allowed model requires a localized interaction on the IR brane of the PQ throat.    In section 4 we then, following Choi\cite{Choi}, turned to a discussion of the realization of the axion via a bulk $U(1)$ gauge symmetry broken by boundary conditions.  Again a two throat model can successfully solve both the $f_{PQ}$ scale problem and the hierarchy problem, but only if QCD
(and possibly the other SM gauge fields) propagate(s) in the PQ throat as well as the $U(1)_{PQ}$ gauge
field.  Finally we discussed the links between these constructions and earlier composite axion models implied
by the holographic correspondence.  Our 5-dimensional models provide a concrete -- and calculable --
realization of the idea of a composite 4d axion.

\vskip 0.5in
\begin{center}
{\bf Acknowledgments}
\end{center}
\vskip0.05in
The work of TF is supported by ``Evangelisches Studienwerk Villigst e.V." and PPARC Grant No. PPA/S/S/2002/03540A. 
JMR thanks the CERN Theory Group and the Galileo Galilei Institute for Theoretical
Physics, Florence for their hospitality and the INFN for partial support during this work. DM wishes to acknowledge the support of the Natural Science and Engineering Research Council of Canada and the Canada-United Kingdom Millennium Research Fellowship.  This work was also partially supported by the EC 6th Framework Programme MRTN-CT-2004-503369.

\end{document}